\documentclass[12pt]{iopart}
\usepackage[T1]{fontenc}

\makeatletter
\usepackage{pxfonts}

\makeatother

\begin{document}

\title{Minkowski momentum of an MHD wave}

\author{Tadas K Nakamura}

\address{CFAAS, Fukui Prefectural University, 910-1195 Fukui, Japan} 
\begin{abstract}
The momentum of an MHD wave has been examined from the view point
of the electromagnetic momentum expression derived by Minkowski. Basic
calculations show that the Minkowski momentum is the sum of electromagnetic
momentum and the momentum of the medium, as proposed in some of the
past literature. The result has been explicitly confirmed by an example
of an MHD wave, whose dynamics can be easily and precisely calculated
from basic equations. The example of MHD wave also demonstrates the
possiblility to construct a symmetric energy-momentum tensor based
on the Minkowski momentum.
\end{abstract}

\maketitle

\section{Introduction}

The Minkowski-Abraham controversy has been discussed by a number of
authors over a hundred years. Minkowski \cite{minkowski10} proposed
the electromagnetic momentum density in a dielectric medium must be
$\mathbf{D}\times\mathbf{B}$, and Abraham \cite{abraham1909,abraham1910}
proposed $\mathbf{E}\times\mathbf{H}$ for that (in the present letter
symbols have conventional meanings, e.g., $\mathbf{E}$ = electric
field, unless otherwise stated). There have been published numerous
papers on this problem both theoretically and experimentally, but
the final conclusion is still yet to come; papers are still beeing
published in this century (see, e.g., \cite{pfeifer2007colloquium}
for a review).

Several authors \cite{feigel2004quantum,pfeifer2007colloquium,Pfeifer2009a}
pointed that the electromagnetic field inevitably affect the dynamics
of the medium to change its energy-momentum, and therefore, the energy-momentum
of an electromagnetic wave must include the contribution of the medium.
In the present letter we show that the Minkowski momentum is the sum
of electromagnetic momentum and the momentum of the medium. Feigel
\cite{feigel2004quantum} obtained a similar result based on the
Noether's theorem using Lagrangian formulation. Compared to his elegant
approach, the calculation here is rather a down-to-eath type, which
is more closer to the Minkowski's original derivation. This approach
is less elegant, however, easier to understand its meaning intuitively.

Perhaps the largest weak point of the Minkowski momentum is the fact
that the four dimensional energy-momentum tensor does not become symmetric
with this momentum, which means the violation of angular momentum
conservation (see, e.g. \cite{jackson1962classical}). Most of the
past literature argued the legitimacy of the momentum part of the
tensor in this point. Here, in contrast, we elucidate the possibility
to alter the energy part to make a symmetric tensor; provided the
momentum part of the Minkowski energy-momentum tensor includes the
momentum of medium, the same should be true for the energy part. To
treat it in a relativistically consistent way, the mass flux must
be included in the energy flux even in the non-relativistic regime.
The energy-momentum tensor with Minkowski momentum can become symmetric
when the mass flux is taken into account.

The consideration stated above is confirmed by an example of an MHD
wave in a collisionless magnetized plasma. Usually the behavior of
an ordinary medium is complicated and need to calculate microscopic
states of molecules, which is difficult to solve exactly. A collisionless
plasma is, in contrast, easy to calculate its response to the electromagnetic
field from the classical basic equations (Maxwell equations and Newtonian
mechanics). Here in this short letter we use the MHD approximation,
however, if one wishes it is possible to derive an exact solution
of the basic equation system to confirm the result. The result agree
with the {}``frozen-in'' of a magnetized plasma, which has been
confirmed by a wide variety of experimental and observational facts.

\section{Basics}

Microscopic Ampere's equation in a medium is \begin{equation}
-\varepsilon_{0}\frac{\partial\mathbf{E}}{\partial t}+\mu_{0}^{-1}\nabla\times\mathbf{B}=\mathbf{J}\label{eq:ampere}\end{equation}
Suppose there is no external current, and $\mathbf{J}$ consists of
the polarization current $\mathbf{J}_{P}$ and magnetization current
$\mathbf{J}_{M}$, which are generated in response to the electric
field $\mathbf{E}$ and magnetic field $\mathbf{B}$ respectively.
We introduce the polarization vector $\mathbf{P}$ and magnetization
vector $\mathbf{M}$ such that \begin{equation}
\frac{\partial}{\partial t}\mathbf{P=\mathbf{J}}_{P}\,,\;\;\nabla\times\mathbf{M}=\mathbf{J}_{M}\label{eq:defPM}\end{equation}
In this context, $\mathbf{P}$ and $\mathbf{M}$ should be understood
as convenient mathematical expressions to represent the response of
the medium to the electromagnetic field, rather than real physical
entities.

When averaged over a microscopically large but macroscopically small
volume, $\bar{\mathbf{P}}$ and $\bar{\mathbf{M}}$ are assumed to
have simple linear relations to the electromagnetic fields as\begin{equation}
\bar{\mathbf{P}}=\chi_{P}\bar{\mathbf{E}}\,,\;\;\bar{\mathbf{M}}=\chi_{M}\bar{\mathbf{B}}\,.\end{equation}
The linear coefficients $\chi_{P}$ and $\chi_{M}$ are matrices in
general because the medium may not be isotropic (as in our example
of magnetized plasmas). The fields $\bar{\mathbf{D}}$ and $\bar{\mathbf{H}}$
are then defined as macroscopic quantities as \begin{equation}
\bar{\mathbf{D}}=\varepsilon_{0}\bar{\mathbf{E}}+\bar{\mathbf{P}}=\varepsilon\bar{\mathbf{E}}\,,\;\;\bar{\mathbf{H}}=\mu_{0}^{-1}\bar{\mathbf{B}}+\bar{\mathbf{M}}=\mu^{-1}\bar{\mathbf{B}}\label{eq:dielectric}\end{equation}

\section{Minkowski Momentum}

The momentum of microscopic electromagnetic field is $\varepsilon_{0}\mathbf{E}\times\mathbf{B}$
and its conservation law is \begin{equation}
\frac{\partial}{\partial t}(\varepsilon_{0}\mathbf{E}\times\mathbf{B})+\nabla\cdot T+(\mathbf{J}_{P}+\mathbf{J}_{M})\times\mathbf{B}+(Q_{P}+Q_{M})\mathbf{E}=0\,,\label{eq:momentum}\end{equation}
where $T$ is the Maxwell stress tensor and we denote $\partial T_{ij}/\partial x_{i}=(\nabla\cdot T)_{j}$
in short. The polarization/magnetization charge $Q_{P}$ and $Q_{M}$
are the result of polarization/magnetization current ($\partial Q_{P,M}/\partial t=\nabla\mathbf{J}_{P,M}$).
The charge due to magnetization current vanishes when averaged, $\bar{Q}_{M}=0$
since $\bar{\mathbf{J}}_{M}$ satisfies (\ref{eq:defPM}). 

The third and fourth term of (\ref{eq:momentum}) are the Lorentz
and Coulomb force acting on the medium, and therefore, it can expressed
by the momentum change of the medium.\begin{equation}
(\mathbf{J}_{p}+\mathbf{J}_{M})\times\mathbf{B}+(Q_{P}+Q_{M})\mathbf{E}=\frac{\partial}{\partial t}\mathbf{g}+\nabla\cdot T_{M}\,,\label{eq:medium}\end{equation}
where $\mathbf{g}$ and $T_{M}$ are the momentum density and stress
tensor of the medium. It should be noted that the right hand side
of the above expression has mathematical ambiguity. If we define new
values of momentum/stress by $\mathbf{g}'=\mathbf{g}+\mathbf{a}$
and $T'=T+G$ with arbitrary vector $\mathbf{a}$ and tensor \textbf{$G$
}that satisfy $\partial\mathbf{a}/\partial t=\nabla G=0$, they also
satisfy the above equation. Therefore, $\mathbf{g}$ and \textbf{$T$}
do not necessarily have to be the total momentum/stress of the medium.
For example, the medium may contain a part that does not interact
with the electromagnetic field, and such part causes this ambiguity.

From (\ref{eq:dielectric}) we obtain\begin{eqnarray}
\bar{\mathbf{J}}_{P}\times\bar{\mathbf{B}} & = & \frac{\partial}{\partial t}(\bar{\mathbf{P}}\times\bar{\mathbf{B}})+\chi_{P}\left[(\bar{\mathbf{E}}\cdot\nabla)\bar{\mathbf{E}}+\frac{1}{2}\nabla\bar{\mathbf{E}}^{2}+\bar{\mathbf{E}}(\nabla\bar{\mathbf{E}})\right]\,,\label{eq:PcrosB}\\
\bar{\mathbf{J}}_{M}\times\bar{\mathbf{B}} & = & \chi_{M}\left[(\bar{\mathbf{B}}\bigtriangledown)\bar{\mathbf{B}}+\frac{1}{2}\nabla\bar{\mathbf{B}}^{2}\right]\,.\nonumber \end{eqnarray}
Here we neglected cross terms of fluctuation in averaging as usually
done in this kind of calculation, e.g., $\overline{\mathbf{P}\times\mathbf{B}}=\bar{\mathbf{P}}\times\bar{\mathbf{B}}$.
Combining (\ref{eq:momentum}), (\ref{eq:medium}) and (\ref{eq:PcrosB})
we obtain\begin{equation}
\frac{\partial}{\partial t}(\varepsilon_{0}\bar{\mathbf{E}}\times\bar{\mathbf{B}}+\bar{\mathbf{g}})+\nabla\cdot(\bar{T}+\bar{T}_{M})=\frac{\partial}{\partial t}(\bar{\mathbf{D}}\times\bar{\mathbf{B}})+\nabla\cdot\bar{T}'=0\,,\label{eq:minkowski}\end{equation}
where $\bar{T}'$ is the stress tensor in a dielectric medium defined
as\begin{equation}
\bar{T}'_{ij}=\bar{E}_{i}\bar{D}_{j}+\mu_{0}^{-1}\bar{B}_{i}\bar{H}_{j}-\frac{1}{2}\delta_{ij}(\bar{\mathbf{E}}\bar{\cdot\mathbf{D}}+\bar{\mathbf{B}}\cdot\bar{\mathbf{H}})\,,\end{equation}
which is the sum of the fluxes of electromagnetic momentum and momentum
carried by the medium. 

Now that we understand the Minkowski momentum includes the part of
the medium then so should be for the energy. The energy is equivalent
to mass in relativity, thus the energy flux of the medium must include
mass, and then the flux may take the form of $\bar{\mathbf{D}}\times\bar{\mathbf{B}}$
to make the energy momentum symmetric. We will check this for the
case of an Alfven wave in the following.

\section{MHD wave}

Let us confirm the above discussion with an example of an MHD wave.
Suppose a linearly polarized one dimensional ($\partial/\partial x=\partial/\partial y=0$)
MHD wave (Alfven wave in this case) propagating in the $z$ direction,
which is the direction of the background magnetic field: $B_{0}=B_{Z}$.
The wave amplitude is small enough for linear approximation, and the
plasma velocity is low enough for non-relativistic approximation.
Also we assume {}``cold plasma limit'', which means the thermal
energy of plasma particles is negligibly small. 

The wave has an electric field perpendicular to its propagation direction,
and we take the $x$ axis in this electric field direction. Then the
current is also in the $x$ direction, whereas the magnetic perturbation
and the plasma velocity is in the $y$ direction (see Appendix). The
magnetic field created by cyclotron motion of plasma particles is
negligible in a cold plasma limit, and thus we treat the case with
$\mu=\mu_{0}$ hereafter.

The current is the polarization current due to the temporal change
of the electric field, which is\begin{equation}
J_{z}=\frac{\mu_{0}}{V_{A}^{2}}\frac{\partial}{\partial t}E_{x}\label{eq:poldrift}\end{equation}
where $V_{A}$ is th Alfven speed defined by $V_{A}=B_{0}/\sqrt{\mu_{0}\rho}$
with $\rho$ being the mass density of the plasma. From (\ref{eq:defPM})
and the above expression we obtain\begin{equation}
D_{x}=\varepsilon_{0}\left(1+\frac{c^{2}}{V_{A}^{2}}\right)E_{x}\label{eq:alfven1}\end{equation}
The plasma frozen-in condition $\mathbf{E}+\mathbf{v}\times\mathbf{B}=0$
means that the plasma is moving in the $y$ direction with the $E\times B$
drift speed as\begin{equation}
v_{y}=\frac{E_{x}}{B_{0}}\,.\label{eq:ecrossb}\end{equation}
Then the momentum carried by the plasma particles is\begin{equation}
\rho v_{y}=\frac{\mu_{0}B_{0}}{V_{A}^{2}}E_{x}\,.\label{eq:plasmamomentum}\end{equation}
The $y$ component of the Minkowski momentum can be calculated from
(\ref{eq:alfven1}) and (\ref{eq:plasmamomentum}), which is\begin{equation}
(\mathbf{D}\times\mathbf{B})_{y}=D_{x}B{}_{0}=\varepsilon_{0}E_{x}B_{0}+\rho v_{y}\,.\label{eq:DcrossB}\end{equation}
The above expression means the Minkowski momentum is the sum of electromagnetic
momentum and momentum of the plasma particles as long as the frozen-in
condition is satisfied. 

When multiplied by $c^{2}$, the first term of the right hand side
of (\ref{eq:DcrossB}) becomes the electromagnetic energy flux in
the $y$ direction. The second term becomes the relativistic energy
comes from the rest mass, which is the predominant energy flux in
the non-relativistic limit here; thermal or kinetic energy flux is
negligible. The balance equation of the energy-momentum tensor is
in a derivative form, and therefore, it has an ambiguity as discussed
below (\ref{eq:medium}) for the momentum. The same is true for the
energy, and its conservation also holds when we add the mass density
and mass flux ($\rho,\rho\mathbf{v})$, since $\partial\rho/\partial t+\nabla(\rho\mathbf{v})=0$.
The energy momentum tensor becomes symmetric with these terms.

\section{Concluding Remarks}

Momentum carried by an electromagnetic wave has been examined with
an example of an MHD wave. It has been shown that the total momentum
(electromagnetic momentum plus momentum of the medium) is expressed
by $\mathbf{D}\times\mathbf{B}$ as proposed by Minkowski. This result
is based on the very simple and basic two properties of an MHD plasma,
the frozen-in condition (\ref{eq:ecrossb}) and polarization current
(\ref{eq:poldrift}), namely. It would not be exaggeration if one
says the whole kingdom of MHD plasma physics would fall if these two
basic properties were wrong. 

Here in this letter we examined a simplest case of an parallel (to
the $\mathbf{B}$ field) propagating MHD wave, but similar calculations
can be done for more complicated plasma waves to confirm the result
here. A collisionless plasma contains a wide variety of wave phenomena,
and the properties of waves can be precisely calculated at least in
the linear limit. Calculation of the Minkowski momentum for various
plasma waves would be a good exercise to understand the Abraham-Minkowski
controversy.

The drawback of the Minkowski momentum has been believed that the
momentum fails to form a symmetric four dimensional energy-momentum
tensor when coupled with the Poyinting flux; an asymmetric energy-momentum
tensor means the violation of angular momentum conservation. This
difficulty can be overcome when we include the mass flux as a part
of energy flux, which is reasonable from the relativistic point of
view. The energy-momentum tensor can be symmetric as we have examined
with an MHD wave here.

What we have shown in the present letter is that the Minkowski momentum
can be self consistent description of the total momentum of an electromagnetic
wave in a polarizable medium. This does not necessarily mean the Abraham
momentum is wrong and inconsistent; it might be possible to give Abraham
momentum another appropriate meaning to make it consistent. For example,
Barnett \cite{Barnett2010} recently argued both Abraham and Minkowski
momentum can be consistent when we interpret the former as kinetic
momentum and latter as canonical momentum. It is out of our scope
here to examine this argument, however, it should be noted the legitimacy
of the Minkowski momentum does not automatically exclude the validity
of the Abraham momentum.

\section*{Appendix}

This appendix is to derive (\ref{eq:poldrift}) and (\ref{eq:ecrossb})
in a shortest way for a physicist not familiar with plasma physics.
For further information, see any textbook on plasma physics, e.g.,
\cite{chen1976introduction,dendy1990plasma}. Note that many books
derive Alfven waves from the MHD equations, which is different from
the derivation here; of course the result is the same.

Suppose a plasma consists of equal number of protons and electrons
in a uniform magnetic field, which is in the $z$ direction of Cartesian
coordinates. The plasma response to the electromagnetic field can
be expressed by $\mathbf{P}$ only and we do not need the magnetization
current for our calculation. Therefore we can set $\mathbf{H}=\mu_{0}^{-1}\mathbf{B}$
here. We assume an MHD wave described above is propagating in this
plasma.

Let us denote a vector in the $xy$ plane by a complex number as $A=A_{x}+iA_{y}$.
Then the wave electric filed in the $x$ direction is denoted as\begin{equation}
E(t)=\frac{E_{0}}{2}(e^{-i\omega t}+e^{\omega t})\,.\end{equation}
The equation of motion of a plasma particle in the $xy$ plane is
written as\begin{equation}
\frac{dv}{dt}=i\Omega v+\frac{e}{m}E(t)\,,\end{equation}
where $\Omega=eB/m$ is the gyro frequency. We include the sign of
the charge in $\Omega$, thus $\Omega$ is positive/negative for a
proton/electron. The above equation can be directly solved as\begin{equation}
v=v_{0}e^{i\Omega t}+\frac{eE_{0}}{2m}\left(\frac{e^{-i\omega t}}{\Omega-\omega}+\frac{e^{i\omega t}}{\Omega+\omega}\right)\,,\label{eq:gyro}\end{equation}
where $v_{0}$ is the integration constant.

Now we assume the wave frequency is much smaller than the gyro frequency
($\omega\ll\Omega$), which is true for most of MHD waves. Then the
effect of the first term in (\ref{eq:gyro}) will be averaged out
for MHD time scale; we do not pay attention to this term hereafter. 

The rest of the motion is called {}``drift'' in plasma physics.
The drift velocity $v_{d}$ can be expanded as\begin{equation}
v_{d}=\frac{E_{0}}{B}\left(i\cos\omega t+\frac{\omega}{\Omega}\sin\omega t+\cdots\right)\,.\label{eq:drifts}\end{equation}
The first term of (\ref{eq:drifts}) is called $E\times B$ drift;
protons and electrons drift in the same direction with the same speed
with this drift. This term is pure imaginary, which means the drift
is in the $y$ direction. This drift gives the predominant motion
of the bulk plasma as in (\ref{eq:ecrossb}), however, it does not
cause a current because both protons and electrons have the same drift
velocity. What contribute to a current is the second term of (\ref{eq:drifts}),
which is called the polarization drift. Since this term contains $\Omega^{-1}$
factor, protons and electrons moves in the opposite direction with
different speed. The electron gyro frequency is much larger than that
of protons, therefore, protons predominantly carry currents. The drift
direction is the same as the electric field since it is pure real,
and the drift speed is proportional to time derivative of the field
because of the factor $\omega$ and $\cos\omega t\rightarrow\sin\omega t$.
Multiplying the proton's second term of (\ref{eq:drifts}) with the
number density and charge, and replacing the factor $\omega$ and
$\cos\omega t\rightarrow\sin\omega t$ by the time derivative, we
obtain (\ref{eq:drifts}).

From the Maxwell's equation we have\begin{equation}
\nabla\times\nabla\times\mathbf{E}=c^{-2}\partial^{2}\mathbf{E}/\partial^{2}t+\mu_{0}\mathbf{J}\,.\end{equation}
When we assume the wave propagation is in the $z$ direction ($\partial/\partial x=\partial/\partial y=0$)
and use (\ref{eq:poldrift}), we obtain the propagation equation of
an MHD wave (Alfven wave) as\begin{equation}
\left(\frac{1}{c^{2}}+\frac{1}{V_{A}^{2}}\right)\frac{\partial^{2}}{\partial t^{2}}E_{x}-\frac{\partial^{2}}{\partial z^{2}}E_{x}=0\,.\end{equation}
The Alfven speed $V_{A}$ is often much smaller than the speed of
light in space and laboratory plasmas. We obtain an wave propagating
with the Alfven speed $V_{A}$ when we ignore the $1/c^{2}$ term
in the above expression.

\section*{References}
\bibliographystyle{unsrt}
\bibliography{Abraham-Minkowski}

\end{document}